\documentclass[onecolumn,nofiglist,notablist]{rsauthor}

\usepackage{graphicx}
\usepackage{amsmath}
\usepackage{amssymb}
\usepackage{amsfonts}
\usepackage{amsthm}
\usepackage{endfloat}
\usepackage{setspace}
\usepackage{verbatim}
\usepackage{geometry}
\usepackage{times}
\usepackage{helvet}
\usepackage{courier}
\usepackage{bm}
\usepackage{url}
\usepackage{dcolumn}
\usepackage{multirow}
\usepackage{color}
\usepackage{natbib}
\usepackage[normalem]{ulem} 

\jname{Phil. Trans. R. Soc. B}
\begin{document}

\title[Network-guided pattern formation of neural dynamics]{Perspective: Network-guided pattern formation of neural dynamics}

\author[Marc-Thorsten H\"utt, Marcus Kaiser, Claus C. Hilgetag]{
Marc-Thorsten H\"utt\thanks{* Author for correspondence ({m.huett@jacobs-university.de})}$^{1*}$, Marcus Kaiser $^{2,3}$, Claus C. Hilgetag$^{4,5}$}

\address{$^1$ School of Engineering and Science, Jacobs University Bremen, Bremen, Germany\\
$^2$ School of Computing Science, Newcastle University, Claremont Tower, Newcastle upon Tyne NE1 7RU, UK\\  
$^3$ Institute of Neuroscience, Newcastle University, Framlington Place, Newcastle upon Tyne NE2 4HH, UK\\
$^4$ Dept. of Computational Neuroscience, University Medical Center Eppendorf, Hamburg University, Germany\\
$^5$ Department of Health Sciences, Boston University, Boston, USA\\
}

\label{firstpage}

\abstract{ 
The understanding of neural activity patterns is fundamentally linked to an understanding of how the brain's network architecture shapes dynamical processes. Established approaches rely mostly on deviations of a given network from certain classes of random graphs. Hypotheses about the supposed role of prominent topological features (for instance, the roles of modularity, network motifs, or hierarchical network organization) are derived from these deviations. An alternative strategy could be to study deviations of network architectures from regular graphs (rings, lattices) and consider the implications of such deviations for self-organized dynamic patterns on the network.  
Following this strategy, we draw on the theory of spatiotemporal pattern formation and propose a novel perspective for analyzing dynamics on networks, by evaluating how the self-organized dynamics are confined by network architecture to a small set of permissible collective states. In particular, we discuss the role of prominent topological features of brain connectivity, such as hubs, modules and hierarchy, in shaping activity patterns. We illustrate the notion of network-guided pattern formation with numerical simulations and outline how it can facilitate the understanding of neural dynamics.}


\maketitle

\section*{Background: Self-organized dynamic patterns in complex brain networks}

A wide range of biological systems are organized in a network-like fashion. Accordingly, the large and diverse field of \emph{network science} has since its very beginning resorted to biological examples to motivate, propose and refine methods for the analysis of complex networks (see for example \citet{Strogatz:2001p116,Barabasi:2004p17908,Barabasi:2012ca}). 
In this way, network science has become a new important paradigm for the understanding of biological systems. Clearly, one of the most fascinating examples of a biological network is the brain. The way in which the brain's network topology shapes, organizes and constrains dynamical processes has received a great amount of attention in recent years and has provided new perspectives in theoretical neuroscience \citep{Bullmore:2009iv,Sporns:2011NotB}.

Another diversely explored paradigm for the understanding of biological systems is the concept of \emph{self-organized} patterns, where collective modes of the system emerge from the local interactions of components (see for example \citet{mikhailov2002}). Diverse forms of distributed computation and global organization are implemented in biological systems via such local interactions,
from the rich ornaments of sea shells and the diversity of animal coat patterns to the myriad of fractal structures in biology and pattern-forming colonies of bacteria. Particularly fascinating are patterns changing with time, resulting in \textit{spatiotemporal} patterns, such as propagating waves and aggregation streams. Bacteria form large branched and nested aggregation-like patterns to immobilize themselves against water flow (see \citet{Levine:2004bk} for a review of various such forms of pattern formation). The individual amoeba in \textit{Dictyostelium discoideum} colonies initiates a transition to a collective multicellular state via a quorum-sensing form of communication: a cAMP signal propagating through the community in the form of spiral waves and the subsequent chemotactic response of the cells lead to branch-like aggregation streams (see, for example, \citet{Kessler:1993ur,Sawai:2005ua,Geberth:2009cm}).  So far, however, these patterns have been mostly discussed for regular interaction architectures, such as lattices.

In this review, we explore a novel view where these two paradigms, network science on the one hand and self-organized pattern formation on the other, are functionally integrated. We discuss some recent findings regarding dynamical processes in topologically complex brain networks, to demonstrate the occurrence of pattern formation guided by the characteristic network architecture. Furthermore, we illustrate with a few simple examples that network-guided pattern formation is a universal and unifying approach for understanding a heterogeneous set of observations about neural dynamics in structured graphs.
Our goal, thus, is to provide the first steps in a unifying framework for these diverse perspectives, explaining how dynamics and topology are tuned in a synergistic fashion via network-guided biological self-organization. 

We focus on the organization of excitable dynamics on graphs. On regular graphs (i.e., rings, lattices), the natural approach of describing dynamical processes is by resorting to the language of large-scale spatiotemporal patterns emerging from local interactions in a self-organized fashion. The exact layout of the patterns is typically determined by random fluctuations or by systematic differences between the nodes of the graph. Here we show that on a graph with less regularity, patterns can be confined by the network architecture to a few network-compatible modes. This phenomenon of network-guided pattern formation can facilitate the interpretation of neural dynamics. 

The logic of this paper is as follows. First, we describe some fundamental topological features of brain networks that have received attention over the last few years, in particular, their heterogeneous degree distribution, resulting in the existence of hubs; as well as modules and a hierarchical organization of networks. 
Next we introduce two minimal dynamical models, helping us to probe these dynamics for the phenomenon of network-guided pattern formation, specifically, reaction-diffusion dynamics as the prototype of pattern-forming dynamical systems, and a simple three-state model of excitable dynamics, which has been employed in various systems for studying the interplay of network topology and dynamics. 
Finally, we attempt to derive from these observations some tentative general conclusions for the organization of brain dynamics.

\subsection*{Theories of spatiotemporal pattern formation}
Theories of spatiotemporal pattern formation have contributed fundamentally to a deep understanding of natural processes, particularly in biology.
One striking example is Turing\textquoteright{}s concept of reaction-diffusion processes, which has a vast range of applications \textendash{} from biology to social systems \citep{Turing1952}. At the same time, these theories (or classes of models) are well embedded in the broader framework of self-organization. 

Self-organization is the emergence of large-scale patterns, based on collective dynamical states, from local interactions. 
Clearly, on regular architectures (like rings or lattices) the emergence of patterns can be easily assessed. In more intricately connected systems, such collective states have been described only for simple cases such as synchronization \citep{Arenas:2006ba}. Over the last years, some progress has been made in extending the concept of patterns on graphs to more general forms of dynamics, for example, to reaction-diffusion systems \citep{Nakao:2010ul} and to a wave-like organization of excitable dynamics around hubs \citep{MullerLinow:2008ia}.

\subsection*{Merging the perspectives of pattern formation and complex networks}

Very much in the light of \citet{Nakao:2010ul} and \citet{MullerLinow:2008ia}, we want to understand what the network equivalents of classical spatiotemporal patterns are, and how, for example, the presence of hubs and modules in networks relates the processes behind spatiotemporal patterns to the theory of complex systems. 

In \citet{MullerLinow:2008ia} it was shown that different topological features of complex networks, such as node centrality and modularity, organize the synchronized network function at different levels of spontaneous activity. Essentially two types of correlations between network topology and dynamics were observed: waves propagating from central nodes and module-based synchronization. These two dynamic regimes represent a graph-equivalent to classical spatiotemporal pattern formation. Remarkably, the dynamic behavior of hierarchical modular networks can switch from one of these modes to the other as levels of spontaneous network activation change. 

In addition, several other studies have attempted to relate notions of spatiotemporal pattern formation with dynamics on graphs.
\citet{Wang2010} emphasized that a certain form of noise-induced pattern formation, spatial coherence resonance, is suppressed by the presence of long-ranging shortcuts and, in general, a small-world network architecture. \citet{Liao2011} rediscovered the target waves around hub nodes previously described in \citet{MullerLinow:2008ia}. They emphasized that large portions of the graph can be enslaved by such patterns (see also \citet{Qian:2010bh,Garcia:2012ey}). The interesting phenomenon of synchronization waves described by \citet{Levya2011} resorts to an embedding of the network in real space. Synchronization waves in this context are characterized by the degree of information transmission.

The waves-to-sync transition in hierarchical graphs (concentric waves around hubs are gradually substituted by synchronous activity within modules) with an increasing rate of spontaneous activity (as described by \citet{MullerLinow:2008ia}) is one example of such collective modes selected for and stabilized by the graph\textquoteright{}s topology and dynamical parameters. The dominant (and functionally important) feature of hierarchical graphs is that hierarchy (independently of its exact definition) shapes every topological scale. Other graph properties (such as modularity or a broad degree distribution) typically reside on a single scale. Therefore, potentially a large number of self-organized, collective modes can \textquoteright{}lock to\textquoteright{} hierarchical topologies. We argue that this \textquoteright{}versatility\textquoteright{} of hierarchical networks is the main reason for their ubiquity in biological systems.

\subsection*{Essential aspects of the organization of brain networks}
\begin{figure}[htbp]
\begin{center}
\includegraphics[width=\textwidth]{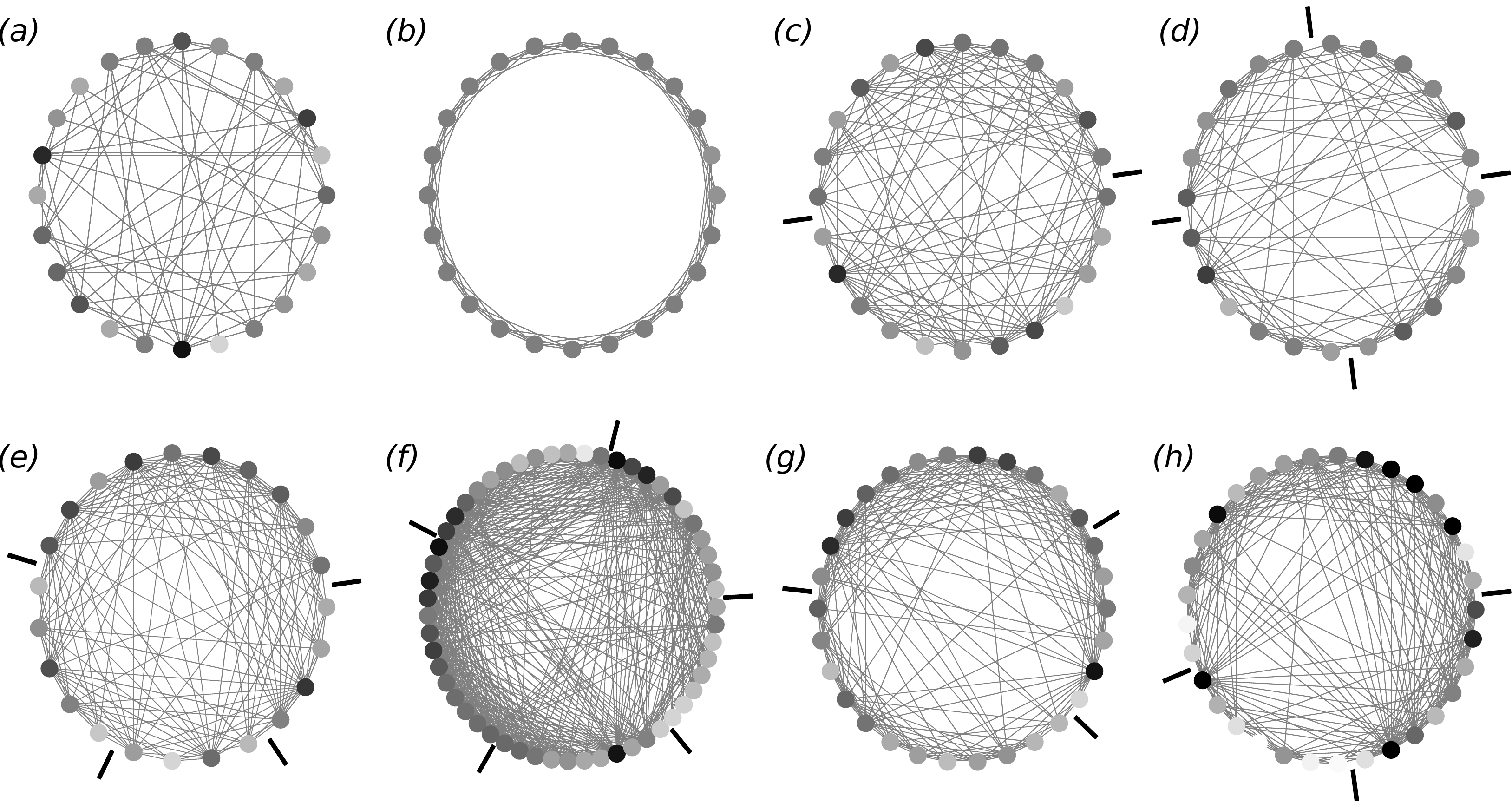}
\caption{Prototype network topologies and brain connectivity examples. (a) Erd\"os-R\'{e}nyi random network. (b) Regular or lattice network with dense connectivity between neighbors. (c) Modular network with two modules. (d) Hierarchical modular network with two modules consisting of two sub-modules each. (e) Rat thalamocortical network of 23 brain regions \citep{Burns2000}. (f) Cat connectivity among 55 cortical and subcortical regions \citep{Scannell1995,Scannell1999}. (g) Connectivity among 30 regions of the primate (macaque monkey) visual cortex \citep{Felleman1991,Hilgetag2000b}. (h) Connectivity among 33 human brain regions (left hemisphere) based on diffusion spectrum imaging \citep{Hagmann2008}. In all panels, regional nodes are arranged on a circle, with node color indicating the degree of the node, that is, the number of its connections (light gray: low degree; dark gray: high degree). Nodes are arranged as to minimize the step distance along the circle between connected nodes, thus also indicating regions of densely connected network modules (module borders are indicated through bars outside the circle).}
\label{archetypes}
\end{center}
\end{figure}

Brain networks can show features of different prototype networks (Figure~\ref{archetypes}). For example, an individual brain network might possess properties of small-world, modular, or hierarchical networks. Network topology might also differ at different scales of network organization, for instance, showing random or regular axonal connectivity at the scale of small neuronal populations and modular connectivity for cortico-cortical fiber tracts \citep{Kaiser2011Network}.  Generally, however, biological neural networks do not conform completely to any of such benchmark networks. Instead, they combine different topological features, including a non-random degree distribution, the existence of network modules as well as the hierarchical combination of such features at different scales of organization. These features are reviewed in the following paragraphs.

\paragraph*{Heterogenous degree distribution}
The term random network typically refers to Erd\"os-R\'{e}nyi random networks \citep{Erdoes1960}, in which  potential connections between nodes are established with a probability $p$. This probability, for a sufficiently large network, is equivalent to the edge density of the network; that is, the connection density. In the creation of random networks, the process of establishing connections resembles flipping a coin where an edge is established with probability $p$ (and not established with probability $q = 1 - p$). Thus, the distribution of node-degrees follows a binomial probability distribution. Neural networks, however, also contain highly-connected nodes, or \textit{hubs}, that are unlikely to occur in random networks. Examples for such highly connected hubs are subcortical structures, such as the amygdala and hippocampus or cortical structures, such as the frontal eye field (FEF) and the lateral intraparietal region (LIP) \citep{Kaiser2007EJN,Sporns2007}. Therefore, neural systems have a heterogeneous degree distribution containing hubs, and share some of the features of scale-free networks \citep{Eguiluz2005,Kaiser2007EJN}.  

\paragraph*{Modules}
Another near-ubiquitous feature of brain networks is the occurrence of modules, within which network nodes are more frequently or densely linked than with the rest of the network \citep{Hilgetag2000b,Hilgetag2004}.  Modularization may be a consequence of the increasing specialization and complexity of neural connectivity in larger brain networks \citep{Sporns2000}. 
Sensory organs and motor units require functional specialization, which begins with the spatial aggregation of neurons into ganglia or topologically into modules (Figure~\ref{archetypes}C), as in the roundworm \textit{Caenorhabditis elegans} \citep{White1986,Achacoso1992,Cherniak1994}. For the cortical network of the cat (Figure~\ref{archetypes}F), modules correspond to large-scale functional units for fronto-limbic, somatosensory-motor, auditory, and visual processing. Spatial and topological modules do not necessarily overlap \citep{Costa2007}, however both tend to be well connected internally, with fewer connections to the rest of the network. There exists a wide range of different algorithms to detect clusters of a network, for example, \citep{Hilgetag2000b,Girvan2002,Palla2005}.

\paragraph*{Hierarchy}
Another reflection of network complexity is the combination, or encapsulation, of topological features at different scales of network organization, which may be termed hierarchy. For example, small modules may be encapsulated in larger modules, which in turn may be contained in even larger modules, resulting in hierarchical modular networks (Figure~\ref{archetypes}D) \citep{Sporns2006,Kaiser:2007td}. 
One example of such hierarchical modularity is the cortical visual system of the non-human primate, the rhesus macaque monkey. Here the visual module consists of two network components (Figure~\ref{archetypes}G): the dorsal pathway for processing object movement and the ventral pathway for processing objects features such as colour and form \citep{Young1992,Felleman1991,Hilgetag2000b}.   Alternative concepts of network hierarchy exist that are based on a sequential network organization or a local versus global access of network nodes (such as in networks with hubs).

Out of these topological features, hierarchy, though poorly conceptualized at the moment, is particularly interesting. Hierarchical organization is an essential attribute of complex biological networks. It implies across-scales information exchange between local signal processing and global integration \citep{Friston2008}. Moreover, hierarchy is linked to aspects of top-down control, regulation, and efficiency (e.g., \citet{Barabasi:2004xy,Yu:2006qf,Gallos2007}) and can hint on developmental principles at evolutionary and ontogenetic scales. For example, gene duplication and area specialization have been discussed as generators of hierarchical neural systems. While there has been impressive progress in understanding biological systems at each hierarchical level (e.g., modeling of single neurons, neuroimaging of the whole brain), the across-scales organization of these systems (i.e., how properties on one scale imply functional features on another scale) is much less well understood, but see \citet{Breakspear2005}.
In general, hierarchical network features have been rarely analyzed, and are only poorly understood at the moment.

\section*{Observations: Features of dynamic patterns in complex brain networks}
\subsection*{Benchmark networks, brain connectivity examples and model dynamics}
We show four different examples for structural brain networks (Figure~\ref{archetypes}E-H), two of which are analysed below regarding the link between topology and dynamics: first, the rat thalamocortical network of 23 cortical and subcortical regions based on tract tracing studies \citep{Burns2000}; second, the cat brain network of 55 cortical and subcortical regions based on tract tracing \citep{Hilgetag2000b,Hilgetag2004,Sporns2004}; third, the rhesus monkey network of 30 brain regions of the visual cortex, excluding the less well-characterised areas Medial Dorsal Parietal  (MDP) and Medial Intraparietal (MIP), based on tract-tracing \citep{Hilgetag2000b,Felleman1991}; and finally, the  network of 33 human cortical regions in the left hemisphere based on diffusion spectrum imaging (DSI) \citep{Hagmann2008}. For visualisation purposes, all nodes were arranged such that the step distance along a circle was minimized for connected nodes, resulting in the modular grouping of densely interconnected sets of nodes \citep{Hilgetag2004}. 

Patterns arising in these networks are explored with two different simple dynamics, a reaction-diffusion system, and a basic excitable model.

\subsection*{An example of network-shaped self-organized dynamics: Turing patterns on graphs}
Let us start with a thought experiment based upon Turing patterns arising in one-dimensional (1D) reaction-diffusion systems. In particular, let us consider these patterns established on a discretized 1D system, that is, a (closed) chain of elements. 

\begin{figure*}[ht]
\begin{center}
\includegraphics[width=0.9\textwidth]{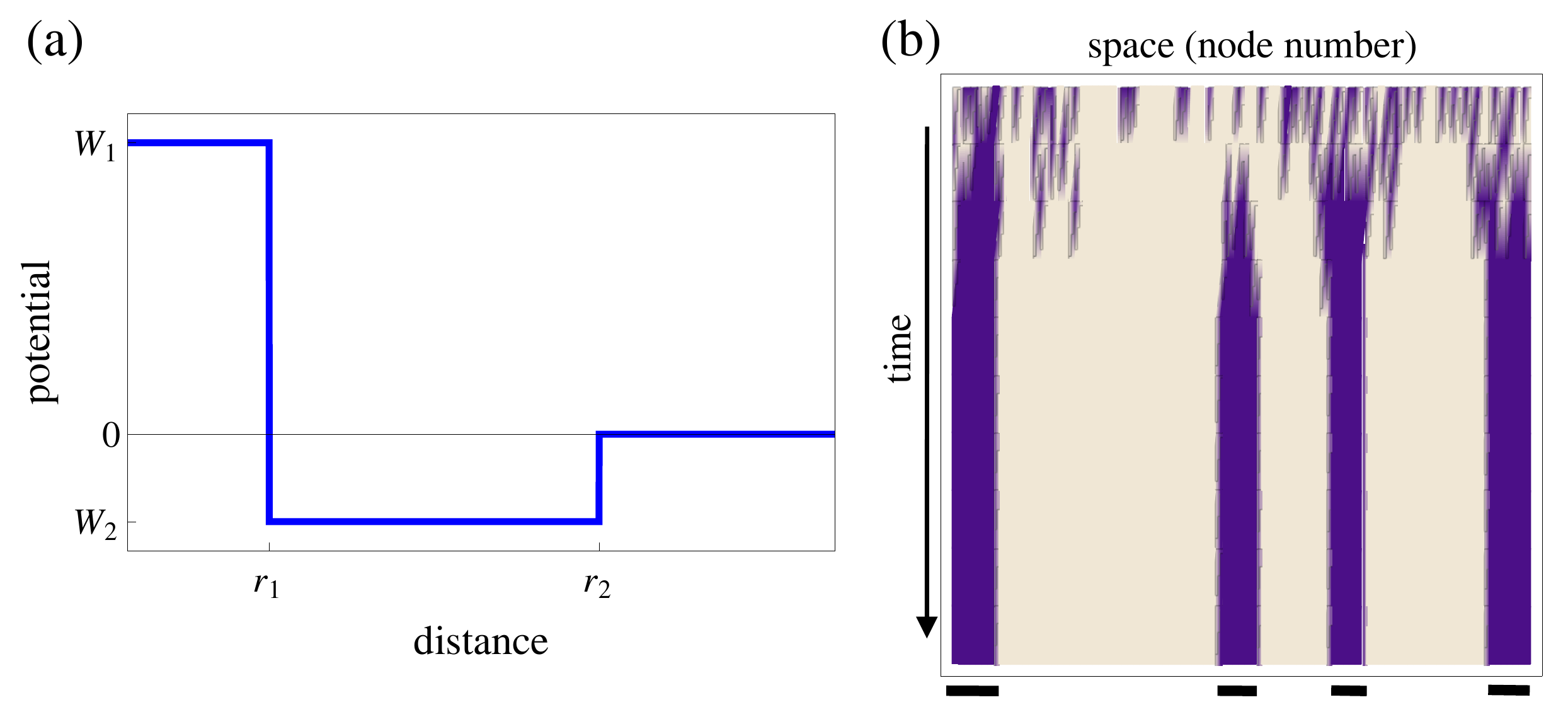}
\caption{Cellular automaton model of a reaction diffusion system: (a) interaction potential of elements as a function of the distance (adapted from \citet{Young:1984tl}); (b) example of a pattern arising in a 1D system (a ring graph); blue: high inhibitor, white: high activator. Parameters are: $r_1=1$, $r_2=3$, $W_1=1$, $W_2=-0.3$, $H=3$.}\label{f1}
\end{center}
\end{figure*}

In order to study such dynamics on arbitrary networks, we here resort to a cellular automaton representation of reaction-diffusion dynamics, similar to the one discussed in \citet{Young:1984tl}. The update rule for each node is given by: 
\begin{equation}
\begin{split}
{x_i}(t + 1) = \Theta \left( H +   {W_1}\sum\limits_{d(i,j) \le {r_1}} {{x_j}(t) + }  {W_2}\sum\limits_{{r_{1 \le }}d(i,j) \le {r_2}} {{x_j}(t)}  \right),
\end{split}
\label{rdca}
\end{equation}
where $d(i,j)$ denotes the (topological) distance between nodes $i$ and $j$ and $\Theta (x) $ yields $1$ for $x>0$ and $-1$ otherwise. 
The quantity $H$ can be considered as an external field biasing the balance of activator ($+1$) and inhibitor ($-1$) states. Figure \ref{f1}(a) sketches the interaction potential underlying the system from eq. (\ref{rdca}), characterized by the range $r_1$ and strength $W_1$ of the activator and the range $r_2$ and strength $W_2$ of the inhibitor. An example of the patterns arising in this system is shown in Figure \ref{f1}(b). Starting from random initial conditions, rapidly a pattern of alternating spatial regions dominated by the activator (white) and the inhibitor (blue), respectively, emerges. 

This is a striking feature of Turing patterns: In spite of the spatial isotropy, some neighboring elements are in identical states, while others display sharp differences. Here, the dynamics self-organize on a spatially homogeneous system (a chain 'network'). 

\begin{figure*}[ht]
\begin{center}
\includegraphics[width=0.9\textwidth]{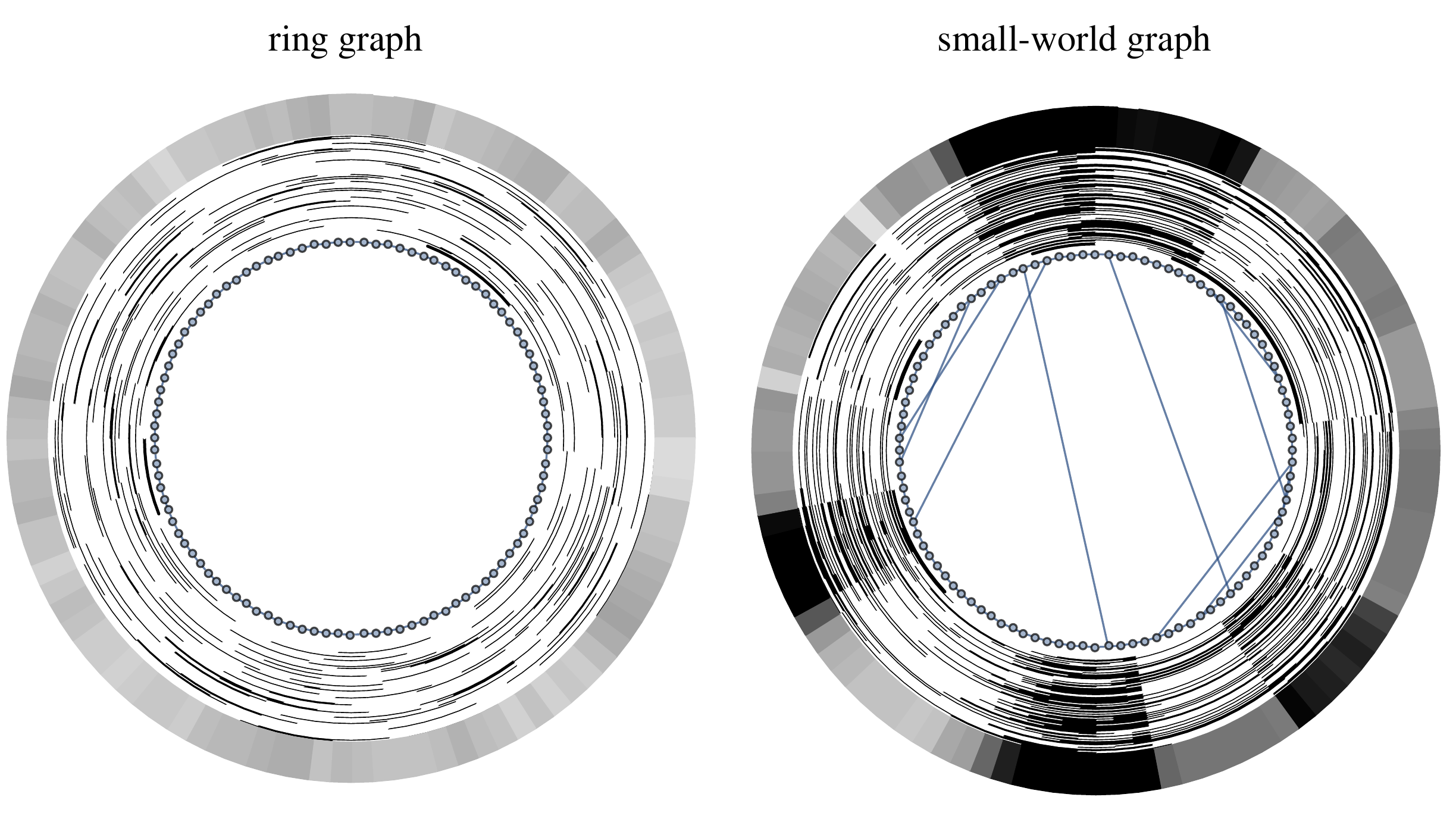}
\caption{On the inside, the network is represented ((a): 100-node ring graph, representing a regular 1D space with periodic boundary conditions; (b): small-world graph obtained from (a) by adding 10 random shortcuts). Around the network the asymptotic high-activator (white) and high-inhibitor (black) regions are shown as rings for 100 runs, each starting from random initial conditions. The outside ring represents the activator-inhibitor asymmetry (number of runs with high activator minus number of runs with high inhibitor computed across the 100 runs shown) for each node. While the patterns average out on the ring graph (a), the shortcuts select certain topology-compatible modes, leading to systematic high-activator and high-inhibitor regions (b). Parameter values are the same as in Figure \ref{f1}. The representation of asymptotic states arranged around the network is the same as indicated below the space-time plot in Figure \ref{f1}b. }\label{f2}
\end{center}
\end{figure*}

The overlay of 100 such asymptotic patterns (Figure \ref{f2}(a)) shows that each spatial site is equally likely to host any of these two regions. 

Let us now disrupt the spatial homogeneity by adding a few long-ranging shortcuts. We see  (Figure \ref{f2}(b)) that the range of possible patterns self-organizing on this systems becomes confined by the spatial inhomogeneities. This is the general phenomenon we would like to call \textit{network-shaped self-organization}.

\begin{figure*}[ht]
\begin{center}
\includegraphics[width=0.9\textwidth]{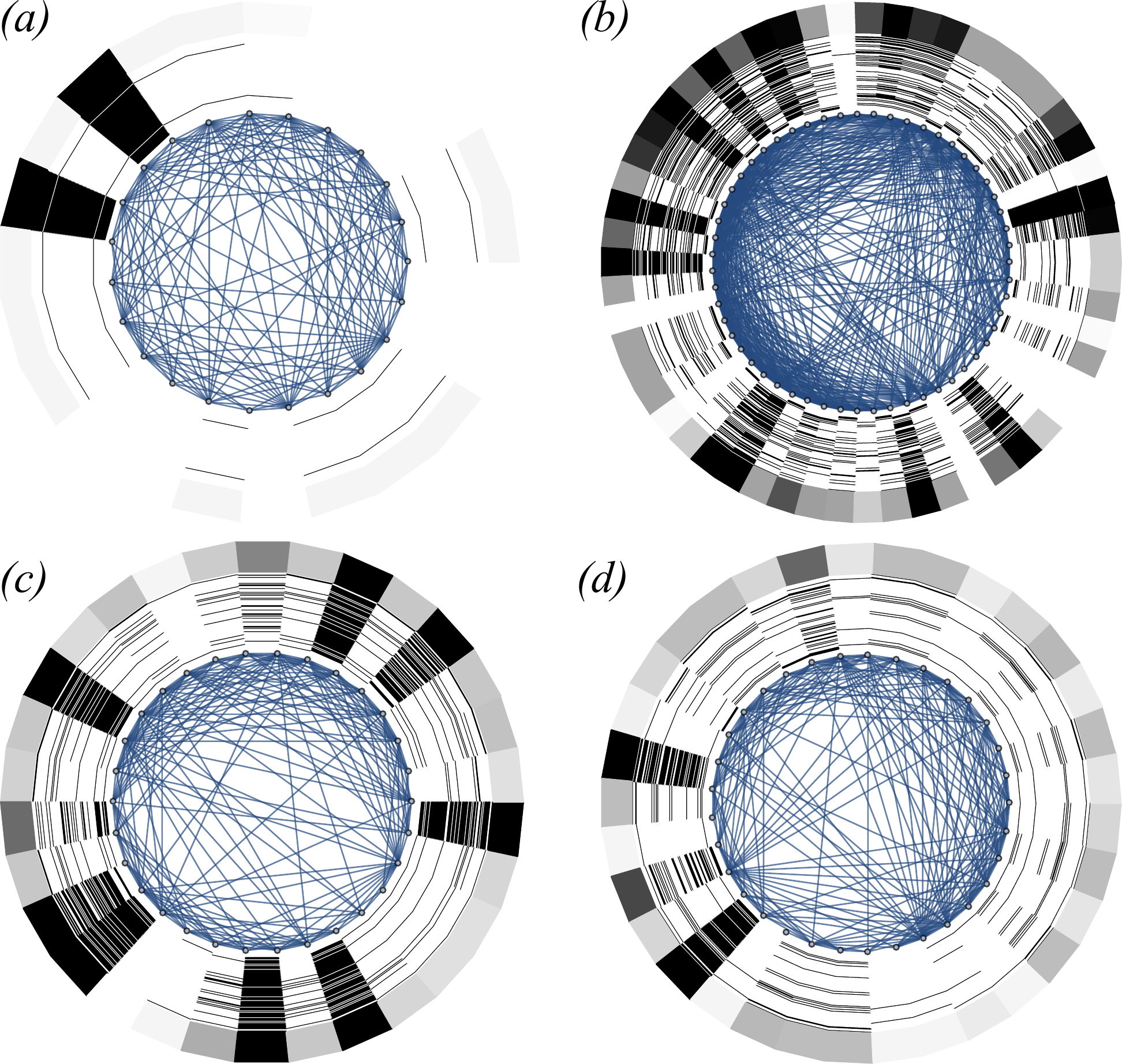}
\caption{ 
Simulation of network dynamics with the same layout as for Figure \ref{f2}, but for the empirical networks shown in Figure \ref{archetypes} (e)--(h): (a) rat thalamocortical network, (b) cat cortical network, (c) macaque visual cortex, (d) human cortical network (left hemisphere).}\label{rdSim}
\end{center}
\end{figure*}

Figure \ref{rdSim} shows the result of activator-inhibitor dynamics (as given by Eq. (\ref{rdca})) on the empirical networks from Figure \ref{archetypes}, that means, for rat, cat, macaque, and human. The two main observations are that (i) the phenomenon of network-guided self-organization is also seen in the network topologies derived from empirical data and (ii) apparently, the confinement of patterns is not trivially linked to select topological features (degree, modularity, etc.), but rather seems to arise from the interplay of several of these features.

\subsection*{Influence of specific topological features on excitable network dynamics}
In a range of previous investigations on excitable dynamics on graphs \citep{MullerLinow:2006ex,MullerLinow:2008ia,Hutt:2009p1867,MullerLinow:2008ia,sr1}, we have identified several examples of  network-shaped self-organization and, in particular, specific topological features serving as 'organizers' of self-organized dynamical modes. 

In the following, we illustrate some of these topological organizers, particularly hubs, modules and network hierarchy. 

We use a three-state cellular automaton model of excitable dynamics, representing a stylized biological neuron or population.
The model has been termed SER model, as each node can be in an susceptible/excitable  ($S$), active/excited ($E$) or refractory ($R$) state. 
The model operates on discrete time and employs the following synchronous update rules: 

A transition from $S$ to $E$ occurs, when at least one neighbor of the $S$ state node is active. After one time step in the state $E$ a node enters the state $R$. The transition from $R$ to $S$ occurs stochastically with the recovery probability $p$, leading to a geometric distribution of refractory times with an average of $1/p$. The model may also include spontaneous transitions from $S$ to $E$ with a probability $f$ (see, e.g.,  \citet{MullerLinow:2006ex,MullerLinow:2008ia,Hutt:2009p1867}). 

In \citet{sr1} a model variant with a relative excitation threshold was used. For a node $i$ with $k_i$ neighbors, the transition from $S$ to $E$ occurs, when at least $\kappa k_i$ neighbors are active. The parameter $\kappa $, thus, serves as a relative excitation threshold. In such a relative-threshold scenario, low-degree nodes are easier to excite (requiring a smaller number of neighboring excitations) than high-degree nodes. 

For $p$=$f$=1, we have a deterministic model, which was investigated in detail in  \citet{Garcia:2012ey}, where the role of cycles in storing excitations and supporting self-sustained activity was elucidated.
The respective influence of hubs (high-degree nodes) and modules in shaping activation patterns has been investigated with a focus on spontaneous excitations \citep{MullerLinow:2008ia,Hutt:2009p1867}. By determining the length of unperturbed propagation of excitations, such spontaneous excitations select the 'topological scale', on which such patterns can be systematically formed. Relatedly, a phenomenon of stochastic resonance (noise-facilitated signal propagation) has been evidenced in so-called 'sub-threshold' networks, that is, for which a single input excitation does not propagate to the output nodes \citep{sr1}. 

The discrete dynamics facilitate a discussion of how excitation patterns are shaped by topological features, due to the possibility of exhaustively mapping all system states and the feasibility of computing large numbers of network and parameter variations.
The approach allowed us to qualitatively assess contributions to functional connectivity and the relationship between structural and functional connectivity.

Let $x_i(t) \in \{S, E, R\}$ be the state of node $i$ at time $t$. It is convenient to discuss the excitation pattern instead: 
\begin{displaymath}
{c_i}(t) = \left\{ {\begin{array}{*{20}{l}}
{ 1, \quad {x_i}(t) = E}\\
{0, \quad {x_i}(t) = S \vee R}
\end{array}} \right.
\end{displaymath}

In this way, we can define a co-activation matrix, 
\begin{displaymath}
{C_{ij}} = \sum\limits_t {{c_i}(t){c_j}(t)} ,
\end{displaymath}
as well as a time-delayed co-activation matrix (or signal propagation matrix),
\begin{displaymath}
{C^{( + )}}_{ij} = {C^{}}_{i \to j} = \sum\limits_t {{c_i}(t){c_j}(t + 1)} 
\end{displaymath}

\begin{figure*}[ht]
\begin{center}
\includegraphics[width=0.6\textwidth]{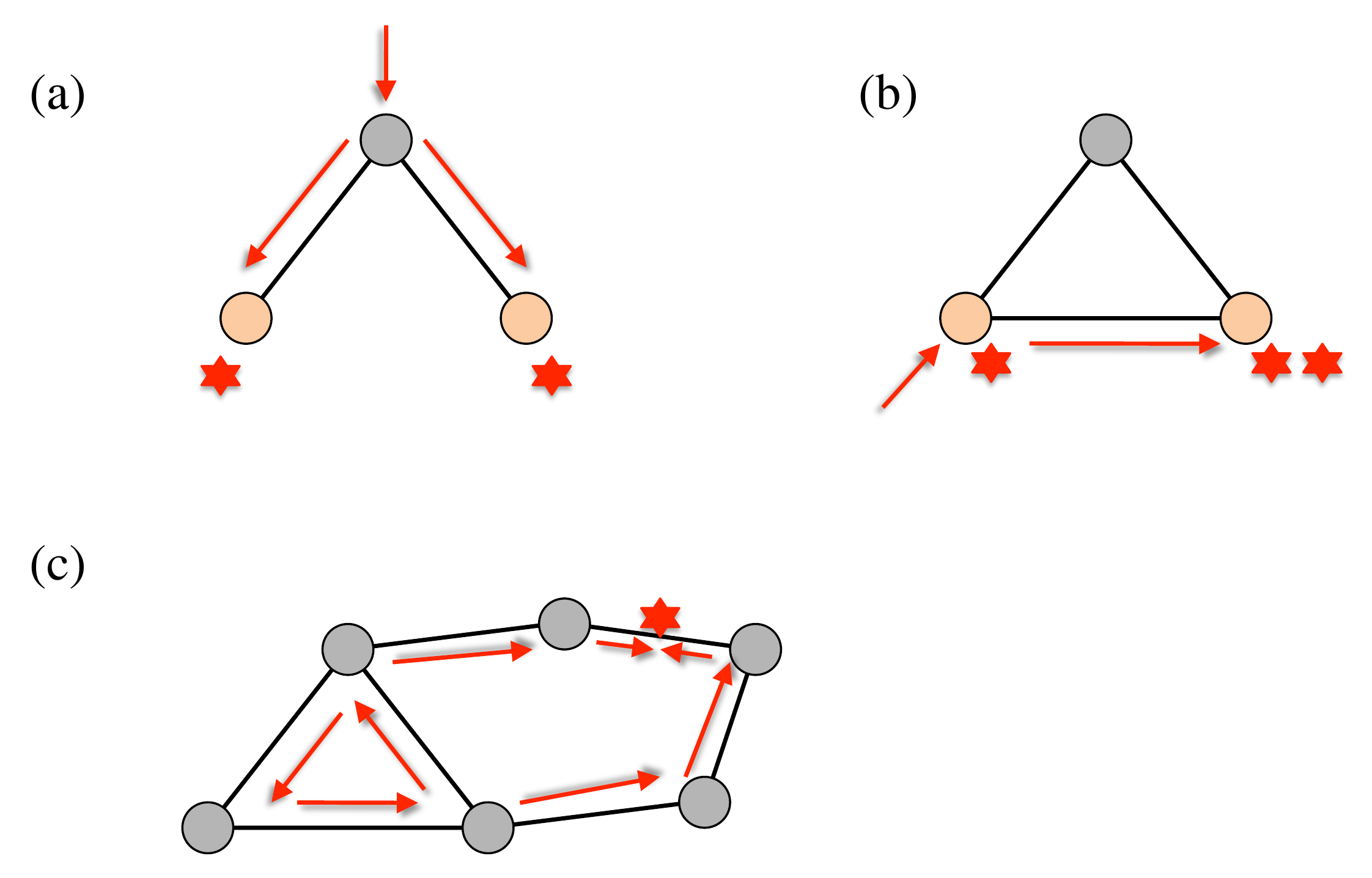}
\caption{Schematic representation of the minimal topological situations underlying co-activation of nodes: (a) co-activation by a common neighbor, (b) sequential activation due to direct links, (c) enslavement of nodes by a short (here: three-node) cycle.}\label{f3}
\end{center}
\end{figure*}

\begin{figure*}[ht]
\begin{center}
\includegraphics[width=0.6\textwidth]{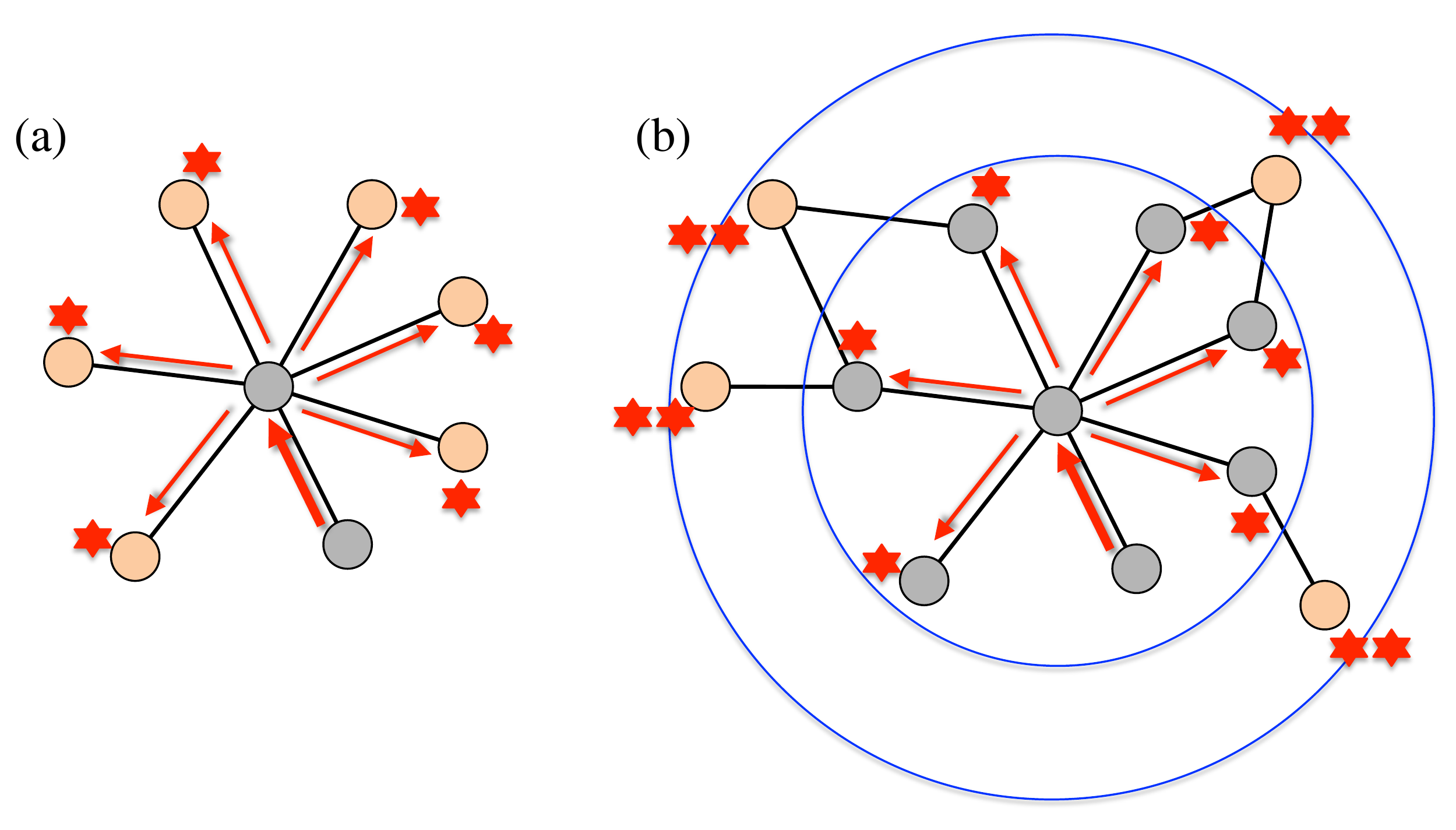}
\caption{Schematic representation of the minimal topological situations leading to ring waves around hubs: (a) An incoming excitation activates the hub and leads to a subsequent excitation of all susceptible nearest neighbors; (b) susceptible nodes with a distance of 2 from the hub are then synchronously activated in the following time step.}\label{f4}
\end{center}
\end{figure*}

Figures \ref{f3} and \ref{f4} compare several minimal topological situations in the context of possible contributions to these matrices. 

Figure \ref{f3}(a) shows a three-chain with an excitation entering at the middle node, leading to a joint excitation at the other two nodes and, consequently,  a contribution to $C_{ij}$. Even though other entry points of excitations, as well as an embedding of this small network 'motif' into a larger network lead to a multitude of other contributions to both, $C_{ij}$ and $C_{i\rightarrow j}$, we can nevertheless deduce that common neighbors lead to an increase in synchronous activity. When a link is added to the two nodes under consideration (thus moving from a three-chain to a three-node loop) the sequential excitation of the two nodes becomes possible (in addition to the previous modes), thus allowing for a contribution to $C_{i\rightarrow j}$. 

Figure \ref{f3}(c) illustrates a more sophisticated contribution to network-shaped self-organization, namely the enslavement of nodes by periodic activity of short cycles. This phenomenon has been analyzed in detail in \citet{Garcia:2012ey}. 

\paragraph*{Heterogeneous degree distribution}
A heterogeneous degree distribution means that some nodes have more connections than others, resulting in the occurrence of hubs, which also have characteristic dynamic features. First, hubs are more active than low-degree nodes. 
Second, \citet{Garcia:2012ey} showed that the node degree is linked to the directed propagation of activity: high-degree nodes (hubs) act as 'senders', whereas low-degree nodes act predominantly as 'receivers' of activity. 
The reasons for this behavior are indicated in Figure \ref{f4}. The figure illustrates the minimal topological mechanism of how propagating waves organize around hubs (as explored in \citet{MullerLinow:2008ia}): Single incoming excitations at a hub are amplified and spread out in a time-synchronous fashion. On a long time scale, the overlay of many such events leads to substantial contributions to $C_{ij}$ in cases where nodes $i$ and $j$ have the same distance from the hub.

\paragraph*{Modules}
In sparsely connected graphs, events of apparent transfer of activity between nodes correspond to actual causal transfers. In such graphs, there may exist a correlation or even anti-correlation between structural links and co-activations, depending on the parameters of the specific dynamic model. For instance, in the simple deterministic SER model outlined above, sparse random graphs show an anti-correlation between structural links and co-activations \citep{Garcia:2012ey}. In denser networks, not all apparent transfers of activity correspond to actual causal transfers. In particular, in dense local neighborhoods of networks, that is, within modules, the local (anti-)correlation between links and co-activations becomes reshaped by the larger-scale network features.  
Specifically, common input of activity from within the same modules results in modular co-activations and appearance of correlation between pairwise links and co-activations \citep{Garcia:2012ey}. This is an important finding, because it suggests that the frequently made observation between structural and functional links in brain connectivity (\citet{Haimovici:2013gk,Honey:2009eh}) is primarily induced by the modular organization of such networks.

\paragraph*{Hierarchy}

Hierarchy can be expressed by different topological features of a network, such as a combination or encapsulation of features, or sequential arrangements of connectivity.  Consequently, there may be different ways in which hierarchy shapes neural dynamics. For example, in hierarchical networks combining modular and hub features, one can observe either hub- or module-driven dynamics of the kind discussed above. These dynamics switch depending on the amount of spontaneous node activation or noise in the system \citet{MullerLinow:2008ia}.  Therefore, this particular hierarchical arrangement provides a transition between different dynamic regimes.

Neural systems are implicitly and explicitly hierarchical. They are
\emph{explicitly hierarchical}, because in many cases the functional components are spread over many scales in space and time (e.g., single ion channels up to brain areas). They are \emph{implicitly hierarchical}, because their organization and underlying interaction patterns (at a specific spatial or temporal scale) often have a nested and layered structure. This implicit hierarchical organization (the network-related hierarchy) has been implicated in a variety of optimal behaviours and dynamic functions by merging different topological features (e.g., modularity and integration). Moreover, hierarchy is related to the compressibility of random walks \citep{Rosvall:2011ey}, to the co-existence of time scales \citep{Arenas:2006ba}, to the range of possible responses upon stimulation \citep{MullerLinow:2008ia} and to the storage of patterns in the networks \citep{Kashtan:2007us}. The impact a hierarchical structure leaves in dynamical processes can qualitatively be described as multi-scale patterns: The distribution of dynamical values across the graph remains invariant under topological coarse-graining (or, more specifically, it obeys well-defined scaling relationships, when such coarse-graining is performed iteratively; \citet{Gallos2007}).

More generally, hierarchical (modular) networks facilitate network-sustained activity \citep{Kaiser:2007td,Kaiser:2010by,Wang:2011gh}, which is a precondition for criticality. The link between network topology and criticality can be made explicit via the topological dimension, which is finite for some (in particular, sparse) hierarchical modular networks, resulting in expanded parameter regimes for criticality, so-called Griffiths phases \cite{Moretti:2013jo}.

\section*{Conclusions}
\subsection*{A new perspective of neural network dynamics}
If the brain were a lattice, neural activity would necessarily produce rich and diverse spatiotemporal patterns, such as spiral waves, synchronous oscillations and concentric waves emanating from periodically firing pacemakers. Noise would be able to interact with the deterministic dynamics to produce coherent activity from, for instance, subthreshold activity. The system would, thus, display noise-facilitated, noise-induced and noise-sustained patterns, according to well-established principles of self-organizing patterns.
Quite obviously, the brain is not a lattice. Neither can a random graph serve as a plausible representation of the intricate overlay of structural elements on all scales observed in real biological neural networks.

Here we have formulated a new perspective on neural dynamics by drawing on concepts of spatiotemporal pattern formation. The heterogeneous network architecture is then viewed as a structural property \textit{confining} patterns to few possible, network-compliant modes. Typical network analyses highlight and investigate deviations from random graphs. Our perspective draws the attention to the deviations from regular graphs, as these deviations are the pattern-confining structural elements.

\subsection*{Specific implications for understanding brain dynamics}

What concrete evidence exists for network-guided self-organization in neurobiological data? 

Based on the topological ingredients discussed above and the subsequent detailed analysis of different dynamical processes on graphs, the notion of network-guided pattern formation points to several building blocks of excitation patterns relevant to neural dynamics and shaped by network topology: 
\begin{enumerate}
\item hubs leading to the propagation of waves \citep{MullerLinow:2008ia},
\item modules leading to localized synchronization, which in turn results in a strong agreement between structural and functional connectivity (as discussed in \citet{Garcia:2012ey}),
\item hierarchical network organization with the potential of facilitating self-sustained activity, criticality, as well as transitions between different dynamical behaviors (see, in particular, \citet{MullerLinow:2008ia} for the latter point). 
\end{enumerate}

Indeed, characteristic spatiotemporal patterns and their implications for functional neural dynamics have been demonstrated in different models of biological neural networks as well as for empirical data. For instance, in the large-scale thalamocortical model of \citet{Izhikevich:2008ti}, the authors describe the emergence of waves and rhythms on different scales.  Additional empirical examples are spreading depression waves associated with retinal migraine \citep{Yu:2012io} and spiral wave dynamics in the neocortex \citep{Huang:2010io}. 

The spatial embedding of the networks can be a principal factor for the arising patterns: activity is spreading to nearby or adjacent patches of neural tissue, in which case networks form a grid or lattice on the brain surface \citep{ODea2013}. Qualitatively speaking, patterns are observed \textit{in spite of} the network, rather than \textit{due to} the network. By contrast, the phenomenon discussed in the present paper addresses the possibility of self-organized patterns where spatial embedding is not the determining factor of the dynamic behavior. In biological terms, connections might link distant brain regions disturbing spatially localised dynamics \citep{Costa2007,Knock2009}. Moreover, these long-distance connections might not even affect delays for activity diffusion due to increased axon diameter or myelination \citep{Buzsaki2013review}. In these cases, the network topology as such dictates the permissible self-organized patterns. 

The most striking example of network-guided self-organization has been discussed in \citet{Moretti:2013jo}, where network heterogeneity generates regions in the network with long activity transients (see also \citet{Hilgetag2013,Munoz:2010gt}). In \citet{Moretti:2013jo}, such Griffiths phase dynamics were suggested as a mechanism for self-sustained activity and critical dynamical states that do not require a careful parameter tuning.
Similar to the Turing patterns arising from reaction-diffusion dynamics and the excitation waves around hubs discussed above, these dynamics are (less regular) forms of collective dynamic behaviors emerging from local interactions.

Criticality is one example of pattern-like self-organized collective dynamics. The importance of critical dynamical states, associated with power-law distributions of activity, have been intensely debated in neuroscience (e.g., \citet{Rubinov:2011ej,Tagliazucchi:2012jx}). 

Remarkably, the network prerequisites discussed for Griffiths phase dynamics and a resulting expanded parameter range for criticality (in particular a specific 'spectral fingerprint' \citep{Moretti:2013jo} which can be directly computed from the adjacency matrix) are similar to the requirements for Turing instability on graphs \citep{Nakao:2010ul} and the synchronizability of graphs \citep{Arenas:2008vg}. 

\begin{figure*}[ht]
\begin{center}
\includegraphics[width=0.6\textwidth]{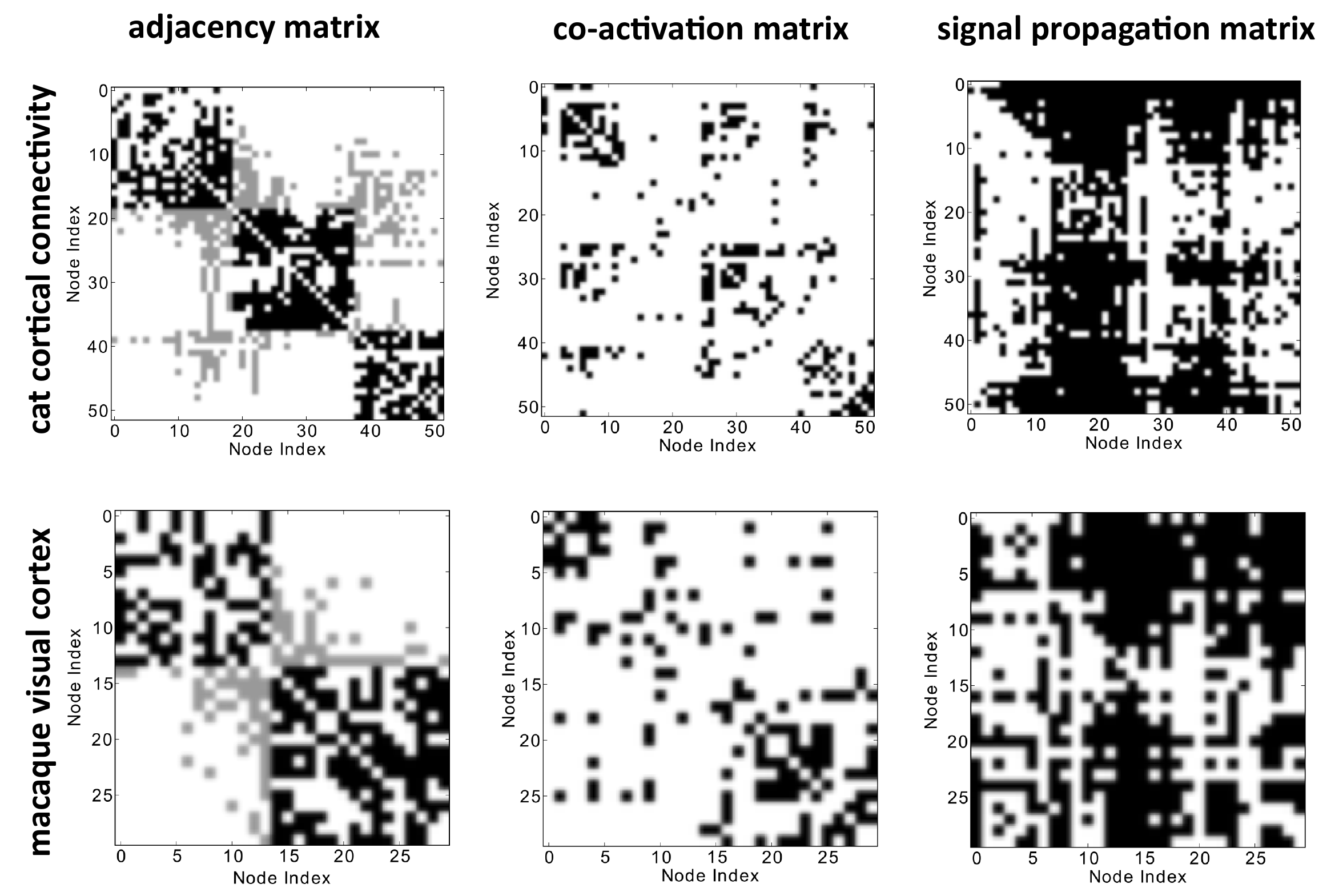}
\caption{Cortical connectivity of the cat (top row) macaque visual cortex (bottom row), together with the resulting activation patterns. 
First column: adjacency matrix (intra-module links are represented in black and inter-module links are represented in gray). Second column: average co-activation matrix $C_{ij}$ binarized with a threshold of 0.46. Third column: time-delayed co-activation (or signal propagation) matrix $C_{i\rightarrow j}$ binarized with a threshold of 0.28. Figure adapted from \citet{Garcia:2012ey}. }\label{coactivation}
\end{center}
\end{figure*}

A series of observations on the agreement of structural and functional connectivity recently established using a simple model of excitable dynamics on graphs \citep{Garcia:2012ey} provides further evidence for network-guided self-organization, in particular, the observation that modules enhance the match between structural and functional connectivity in the (dense) modules, while a broad degree distribution tends to reduce the match due to the organization of activity around hubs. Figure \ref{coactivation}, as well as the discussions in \citet{Garcia:2012ey} provide detailed accounts of these associations. The effects captured by the schematic representations of local dynamics shown in Figures \ref{f4} and \ref{f3} are the underlying microscopic mechanisms for the coactivation patterns observed in Figure  \ref{coactivation}.

Modular node activations, and anti-correlations of different modules, are a prominent and conspicuous feature of functional neural dynamics (e.g., \citet{Fox:2005fr}).  They have been reproduced in a variety of large-scale computational modes (e.g., \citet{Honey2007,Deco:2009rt}).  Our thinking suggests that this phenomenon may primarily result from the spatio-temporal pattern formation in modular neural networks, rather than depend on  particular parameters of the local node dynamics.

As a further example, in  \citet{sr1} it has been observed that signal coherence (measured by the amount of interdependent excitations) is enhanced by noise in a resonant fashion, with noise being  provided by spontaneous excitations. This collective effect is similar to the well-known phenomenon of spatiotemporal stochastic resonance \citep{Jung:1995vz}. 

Finally and generally, the structural ingredients of self-sustained activity have been intensely discussed over the last years (see, e.g., \citet{Deco:2009p6486,Deco:2011p775}). Network-guided self-organization may provide a promising novel framework for better understanding the network requirements for such collective dynamic states of neuronal activity.

\ack{MTH is supported by DFG grant HU 937/7-1. MK was supported by the WCU program through the KOSEF funded by the MEST (R31-10089), EPSRC (EP/K026992/1), and the CARMEN e-science project (\protect{http://www.carmen.org.uk}) funded by EPSRC (EP/E002331/1). CCH is supported by DFG grants HI 1286/5-1 and SFB 936/A1.}

\bibliographystyle{myapalike}
\bibliography{PTRStemp3,neuro}

\end{document}